\documentclass[a4paper,12pt]{article}

\usepackage{amsmath}\usepackage{amssymb}
\usepackage{latexsym}\usepackage{cite}

\usepackage[dvips]{graphicx}
\usepackage{pstricks}
\usepackage{bm}
\usepackage{color}

\usepackage{graphicx}

\definecolor{Red}{rgb}{1.00, 0.00, 0.00}

\newcommand{\be}{\begin{equation}}
\newcommand{\ee}{\end{equation}}

\def\beq{\begin{equation}}
\def\eeq{\end{equation}}
\def\beqr{\begin{eqnarray}}
\def\eeqr{\end{eqnarray}}

\def\al{\alpha}
\def\bt{\beta}

\def\de{\delta}
\def\De{\Delta}

\def\si{\sigma}
\def\Si{\Sigma}
\def\te{\theta}

\def\lam{\lambda}

\def\ep{\epsilon}

\def\sq{\sqrt}

\def\l{\left (}
\def\r{\right )}

\def\fr{\frac}
\def\la{\label}
\def\hs{\hspace}
\def\vs{\vspace}

\def\ran{\rangle}
\def\lan{\langle}
\def\ov{\overline}
\def\tl{\tilde}
\def\tm{\times}

\begin{document}

\begin{flushright}
September 13, 2011 \\
\end{flushright}

\vs{1.5cm}

\begin{center}
{\Large\bf

New Flavor $U(1)_F$ Symmetry for SUSY $SU(5)$}

\end{center}

\vspace{0.5cm}
\begin{center}
{\large
{}~Zurab Tavartkiladze\footnote{E-mail: zurab.tavartkiladze@gmail.com

\hs{0.3cm}Associate member of Andronikashvili Institute of Physics, 0177 Tbilisi, Georgia.}
}
\vspace{0.5cm}

{\em Center for Elementary Particle Physics, ITP, Ilia State University, 0162 Tbilisi, Georgia}

\end{center}
\vspace{0.6cm}

\begin{abstract}


Within supersymmetric $SU(5)$ Grand Unified Theory, we present several new  scenarios with anomaly
free flavor symmetry $U(1)_F$. Within each scenario, a variety of cases offer many possibilities for phenomenologically
interesting model building. We present three concrete and economical models with anomaly free $U(1)_F$ leading to natural
understanding of observed hierarchies between charged fermion masses and CKM mixing angles.

\end{abstract}





\section{Introduction}

Noticeable hierarchies between charged fermion masses and mixings remain unexplained within the Standard Model and its minimal supersymmetric
(SUSY) extension. Grand Unification (GUT) \cite{Pati:1974yy} gives some interesting asymptotic mass relations (like $m_b=m_{\tau }$ within
$SU(5)$ GUT), but problem of flavor still remains unresolved. It may well be that the resolution of this puzzle has some physical origin, and a nice
idea is existence of the flavor symmetry acting between different flavors of quarks and leptons. The simplest possibility is the Abelian $U(1)_F$ flavor symmetry \cite{Froggatt:1978nt},
which has been extensively investigated with $U(1)_F$ being an anomalous symmetry \cite{Ibanez:1994ig} of a stringy
origin \cite{Dine:1987xk}. Some attempts to find anomaly free setup with $U(1)_F$ symmetry, for explanation of fermion mass hierarchies, also
exist in a literature \cite{Dudas:1995yu, Chen:2008tc}. With anomaly free $U(1)_F$, without relying on some specific string construction,
one can investigate a given scenario (within MSSM \cite{Dudas:1995yu} or GUT \cite{Chen:2008tc}) based on conventional field theoretical arguments.

In this Letter within SUSY $SU(5)$ GUT, we suggest new way of finding non-anomalous $U(1)_F$ flavor symmetries. We present several
scenarios with anomaly free $U(1)_F$ symmetries, which provide natural explanation of hierarchies between charged fermion masses and mixings.

The Letter is organized as follows. In the next section we pursue the way of finding non-anomalous $U(1)_F$ symmetries by possible embedding
of $SU(5)\tm U(1)_F$ into the anomaly free non-Abelian symmetry, and present our findings. In section \ref{sect:models}, after listing requirements which we follow upon model building, classify various possibilities within scenarios we have found. Further, we present three concrete models which give natural understanding
of hierarchies between charged fermion Yukawa couplings and CKM mixing angles. Within these models leptonic mixing angle $\te_{\mu \tau }$ turns out to be
naturally large, giving good background for building promising scenarios for neutrino masses and mixings. Our conclusions are given in Sect. \ref{sec:conc},
while in Appendix A we discuss the possibility of consistent $U(1)_F$ symmetry breaking, needed for realistic model building.

\section{SUSY $SU(5)$ and Non-Anomalous Flavor $U(1)_F$}\label{sect:U1s}

As already noted, we are working within the framework of SUSY $SU(5)$ GUT and looking for non-anomalous flavor
$U(1)_F$ symmetry. Minimal chiral content for the fermion sector consists to
$10+\bar 5$ multiplets per generation, whose SM matter composition and quantum numbers under $SU(3)_C\tm SU(2)_L\tm U(1)_Y\equiv G_{321}$ gauge group is
\beq
10=q(3, 2, -\fr{1}{\sq{60}})+u^c(\bar 3, 1, \fr{4}{\sq{60}})+e^c(1, 1, -\fr{6}{\sq{60}})~,~~~
\bar 5=d^c(\bar 3, 1, -\fr{2}{\sq{60}})+l(1, 2, \fr{3}{\sq{60}})~.
\la{su5matter}
\eeq
Last entries in the brackets represent corresponding hypercharge $Y$ with $SU(5)$ normalization (being generator of $SU(5)$, the $Y$ has the form
$Y=\fr{1}{\sq{60}}{\rm Diag}(2, 2, 2, -3, -3)$).
Our aim is to find such family dependent $U(1)_F$ charge assignments which are anomaly free. Clearly, for one of the families,
out of three, the simplest assignment $10_0+\bar 5_0$ (subscripts indicate $U(1)_F$ charges) is anomaly free.
The non-zero charge assignment would require\footnote{Unless $U(1)_F$ charge assignments are such that anomalies coming from different families cancel
each other.} addition of new states which can be $SU(5)$ singlets charged under $U(1)_F$.
We will look for extensions with minimal possible content.
 By minimal content we mean that non-trivial $SU(5)$ representations, which we introduce, will be just those of minimal SUSY $SU(5)$ GUT.
These are three families of matter $(10+\bar{5})$-supermultiplets, one pair of Higgs superfields $H(5)+\bar H(\bar 5)$ (including MSSM Higgs doublet
superfields $h_u$ and $h_d$ respectively), and an adjoint $\Si (24)$ of $SU(5)$ (needed for symmetry breaking $SU(5)\to G_{321}$). We assume that
$\Si $ has no $U(1)_F$ charge, $Q(\Si )=0$, and thus does not contribute to anomalies. Thus $SU(5)$ states which may contribute to anomalies are
three $10_i$-plets ($i=1,2,3$), four $\bar 5_k$-plets ($k=1,2,3,4$) and one $5$-plet. With this set, the $SU(5)^3$ anomaly
$A_{555}=3A(10)+4A(\bar 5)+A(5)=0$ vanishes because $A(10)=-A(\bar 5)=A(5)$. As far as the anomalies $SU(5)^2\cdot U(1)_F$, $\l U(1)_F\r^3$ and
$({\rm Gravity})^2\cdot U(1)_F$ are concerned, they should satisfy the following conditions:
\beqr
SU(5)^2\cdot U(1)_F:& ~~A_{551}&\!\!\!\!=\fr{3}{2}\sum_{i=1}^3Q(10_i)+\fr{1}{2}\l \sum_{k=1}^4Q(\bar 5_k)+Q(5)\r =0~,
\la{A551} \\
&&\nonumber \\
\l U(1)_F\r^3: &  A_{111}&\!\!\!\!=10\sum_{i=1}^3Q(10_i)^3+5\l \sum_{k=1}^4Q(\bar 5_k)^3+Q(5)^3\r +\sum_{s}Q_s^3=0~,
\la{A111}\\
&&\nonumber \\
\l {\rm Gravity}\r^2\cdot U(1)_F: &  A_{GG1}&\!\!\!\!={\rm Tr}Q=10\sum_{i=1}^3Q(10_i)+5\l \sum_{k=1}^4Q(\bar 5_k)+Q(5)\r +\sum_{s}Q_s=0~,
\la{gg1}
\eeqr
where $Q_s$ denotes $U(1)_F$ charges of $SU(5)$ singlet states. Upon finding the anomaly free assignments we will limit ourself with
scenarios involving small number of singlets.
All other mixed anomalies vanish due to properties of  $SU(5)$ generators.

Since $U(1)_F$ is Abelian  symmetry, there is no overall normalization for the charges. However, in order
to have realistic phenomenology, all charge ratios should be rational numbers; i.e. in the unit of one field's charge, all
remaining states' charges should be rational numbers. To find such anomaly free charge assignment, leading to desirable phenomenology,
one way is to find solution(s) of system of Eqs. (\ref{A551})-(\ref{gg1}) in a straightforward way \cite{Dudas:1995yu}, \cite{Chen:2008tc}.
Another way might be to extract $U(1)_F$ (as a subgroup) from anomaly free non-Abelian flavor symmetries \cite{Babu:2011yw} (which are compatible with
$SU(5)$ GUT). Different, and unexplored yet, way of finding is to embed $SU(5)\tm U(1)_F$ (as a subgroup) in higher non-Abelian symmetries with anomaly free
content.
In this work, we follow the latter way in order to find  anomaly free flavor $U(1)_F$ symmetries within SUSY $SU(5)$ GUT.

For this purpose, consider higher gauge
symmetries containing $SU(5)$ as their subgroups plus $U(1)$ factors. Clearly, the rank of such non-Abelian groups should be $\geq 5$.
Since all states will belong to non-Abelian groups, the condition ${\rm Tr}Q=0$ of Eq. (\ref{gg1}) will be automatically satisfied.
However, vanishing of other anomalies will require specific selection of the field content \cite{anomalies}.
One simple possibility emerges via $SO(10)$ group which has a maximal subgroup is
$SU(5)\tm U(1)'$. $SO(10)$'s spinorial representation - the $16$-plet -
decomposes under the $SU(5)\tm U(1)'$ as \cite{Slansky:1981yr}
\beq
16=10_1+\bar 5_{-3}+1_5~,
\la{flipSU5}
\eeq
where subscripts are $U(1)'$ charges
\footnote{We omit normalization factor, which is not essential here.}
which can be
identified with  $U(1)_F$ charges. In this way,  $U(1)_F$ is anomaly free since all anomalies [$SU(5)^3$, $SU(5)^2\cdot  U(1)_F$, etc.] vanish.
The $SU(5)$'s singlet $1_5$, charged under $U(1)_F$, plays important role for anomaly cancellation.

For finding another assignment let us consider $27$-plet of $E_6$ group (the rank six exceptional group). With $E_6\to SO(10)\tm U(1)''\to SU(5)\tm U(1)''$ decomposition we have
\cite{Slansky:1981yr}
\beq
27=16_1+10_{-2}+{1'}_{4}=(10+\bar 5+1)_1+(5+\bar 5')_{-2}+{1'}_{4}~,
\la{flipSO10}
\eeq
where subscripts denote $U(1)''$ charges. In this case $U(1)''$ can be identified with $U(1)_F$. We see that, in this case anomaly cansellation requires two
$SU(5)$ singlets (charged under $U(1)_F$) and extra charged $5, \bar 5$ plets of $SU(5)$.

Each anomaly free content (\ref{flipSU5}) and (\ref{flipSO10}), we presented so far, includes one $10$-plet of $SU(5)$. This happened because of simple and
single anomaly free $SO(10)$ and $E_6$ representations $16$ and $27$-plets respectively. Another (higher) representations might give more $10$-plets.
Since those higher states would also involve extra exotic states, we do not consider such possibilities here.

As far as the unitary groups with rank greater than five, we start discussion with $SU(7)$. Lower group $SU(6)$ is subgroup of $E_6$ which was already
considered above (detailed comment about this is given at the end of this section). As it will turn out,  the $SU(7)$ group can give
an interesting anomaly free field content.
Consider $SU(7)$'s one particular set of chiral representations $35+2\tm \bar 7$, which is anomaly free.\footnote{Other $SU(7)$'s anomaly free chiral sets
like $21+3\tm \bar 7$ and $21+\ov{35}+\bar 7$ etc., involve either too many $SU(5)$ singlets, or unwanted $SU(5)$ states and thus will not be considered here.}
Here $35$ is three index antisymmetric representation and $\bar 7$ is an anti-fundamental of $SU(7)$. Their decomposition via the
chain $SU(7)\to SU(6)\tm U(1)_7\to SU(5)\tm U(1)_6\tm U(1)_7$ is
$$
35=20_3+15_{-4}=(10_{-3}+\ov{10}_3)_3+(10_2+5_{-4})_{-4}~,
$$
\beq
\bar 7=\bar 6_{-1}+1_6=(\bar 5_{-1}+1_5)_{-1}+(1_0)_6~,
\la{SU7}
\eeq
where inside and outside of parenthesis the $U(1)_6$ and $U(1)_7$ charges respectively are indicated as subscripts. Note that $U(1)_6$ and $U(1)_7$ are coming from
$SU(6)$ and $SU(7)$ respectively. Their corresponding generators are $Y_{U(1)_6}=\fr{1}{\sqrt{60}}{\rm Diag}\l 1, 1, 1, 1, 1, -5\r $ and
$Y_{U(1)_7}=\fr{1}{\sqrt{84}}{\rm Diag}\l 1, 1, 1, 1, 1, 1, -6\r $ respectively. The normalization factors $\fr{1}{\sqrt{60}}$   and $\fr{1}{\sqrt{84}}$
are omitted in Eq. (\ref{SU7}). Note that set of Eq. (\ref{SU7}) includes $SU(5)$'s  $\ov{10}$-plet, which we did not intend to introduce.
However, there is one loophole which helps in this situation. Since consideration of $SU(7)$ symmetry was just the way of finding the anomaly free
$U(1)_F$, we will consider $SU(5)\tm U(1)_F$ gauge symmetry, important is that $SU(5)^3$ and anomalies of Eqs. (\ref{A551})-(\ref{gg1})
vanish. So, if the pair of $(\ov{10}+5)$-plets is replaced by $(10+\bar 5)$ then $SU(5)^3$ anomaly will not be changed (i.e. will still vanish). With this
substitution $\ov{10}\to 10$, $5\to \bar 5$, without changing the $U(1)$-charges, the mixed and cubic anomalies  (\ref{A551})-(\ref{gg1})  will remain intact.
Therefore, we will consider the following content
\beq
(10_{-3}+10_3)_3+(10_2+\bar 5_{-4})_{-4}+2\tm \left [(\bar 5_{-1}+1_5)_{-1}+(1_0)_6 \right ]~,
\la{SU7-content}
\eeq
which involves three pairs (three families!) of $(10+\bar 5)$-plets.

Higher rank gauge groups $SU(N>7), SO(N>10), E_7, E_8$ etc. with corresponding anomaly free representations will give extra (unwanted) non-trivial
representations of $SU(5)$
 and we do not consider them. Therefore, we will use the sets (\ref{flipSU5}), (\ref{flipSO10}) and (\ref{SU7-content}) in our further
studies for model building.

With Abelian symmetries $U(1)'$, $U(1)''$, $U(1)_6$ and $U(1)_7$ various linear superpositions can be constructed. Starting with $U(1)'$ and $U(1)''$,
which respectively transform the sets given in Eqs. (\ref{flipSU5}) and (\ref{flipSO10}), let us consider the superposition
\beq
Q_{sup}=aQ_{U(1)''}+bQ_{U(1)'}~.
\la{superposition}
\eeq
In order to construct such a superposition, the content of (\ref{flipSU5}) should be extended with extra singlet $1'$ and $5, \bar 5'$ plets, with
$U(1)'$ charges $Q_{U(1)'}(1')=0$, $Q_{U(1)'}(5)=2q$, $Q_{U(1)'}(\bar 5')=-2q$, where $q$ is some number. $Q_{sup}$ can be identified with $U(1)_F(=U(1)_{sup})$.
In order that $U(1)_F=U(1)_{sup}$ be anomaly free some constraints on $a, b$ and $q$ should be imposed. Simplest possibility, leading to realistic models, is to
require cancellation of mixed anomalies $U(1)'\cdot [U(1)'']^2$ and $[U(1)']^2\cdot U(1)''$. If these mixed anomalies will vanish, then $U(1)_{sup}$ also will be anomaly free (because separately $U(1)'$ and $U(1)''$ are anomaly free). One can easily make sure that with $q=\pm 1$  $U(1)_{sup}$ is anomaly free for arbitrary
values of $a$ and $b$. Note that $q=-1$ corresponds
to $SO(10)$ normalization, i.e.  $SU(5)$'s multiplets $5$ and $\bar 5$ coming from the $SO(10)$'s fundamental $10$-plet, should have charges $-2$ and $2$ resp.
Without loss of generality, we will choose $q=-1$. Thus, anomaly free field content is:
\beq
10_{a+b}+\bar{5}_{a-3b}+1_{a+5b}+5_{-2a-2b}+\bar{5}'_{-2a+2b}+1'_{4a}~.
\la{supS10-E6}
\eeq

Similarly, from $U(1)_6$ and $U(1)_7$ charges of the fields given in Eq. (\ref{SU7-content}) we can build superposition
\beq
\bar Q_{sup}=\bar aQ_{U(1)_6}+\bar bQ_{U(1)_7}~.
\la{superposition1}
\eeq
Note that $\bar Q_{sup}$ is automatically anomaly free for arbitrary $\bar{a}$ and $\bar{b}$, because the orthogonal generators $Y_{U(1)_6}$ and $Y_{U(1)_7}$
originate from single $SU(7)$. Thus, using (\ref{SU7-content}) we can write the anomaly free set
\beq
10_{-3\bar{a}+3\bar{b}}+10_{3\bar{a}+3\bar{b}}+10_{2\bar{a}-4\bar{b}}+\bar 5_{-4\bar{a}-4\bar{b}}+
2\tm \left (\bar 5_{-\bar{a}-\bar{b}}+1_{5\bar{a}-\bar{b}}+1'_{6\bar{b}}  \right )~,
\la{supSU6-SU7}
\eeq
where subscripts denote $\bar Q_{sup}$ charges. These charges could be identified with charges of flavor $U(1)_F$.

Summarizing all possibilities discussed above, we can have the following options for flavor $U(1)_F$ charge assignments:
\begin{eqnarray}
{\bf A}:&& 10_0+\bar{5}_0~, \la{A} \\
{\bf B}:&& 10_{\al }+\bar{5}_{-3\al }+1_{5\al }~,~~(\al \neq 0)~, \la{B} \\
{\bf C}:&& 10_{a+b}+\bar{5}_{a-3b}+1_{a+5b}+5_{-2a-2b}+\bar{5}'_{-2a+2b}+1'_{4a}~,~~(a\neq 0~,~a\neq -5b)~, \la{C} \\
{\bf D}:&& 10_{-3\bar{a}+3\bar{b}}+10_{3\bar{a}+3\bar{b}}+10_{2\bar{a}-4\bar{b}}+\bar 5_{-4\bar{a}-4\bar{b}}+
2\tm \left (\bar 5_{-\bar{a}-\bar{b}}+1_{5\bar{a}-\bar{b}}+1'_{6\bar{b}}  \right ),
~(\bar{b}\neq 0).
\la{D}
\end{eqnarray}
The conditions in brackets are imposed in order to avoid repetition of identical cases. For example, in case {\bf B}, with $\al =0$ we recover
case {\bf A} with extra neutral $SU(5)$ singlet. Likewise, in case {\bf C}, with $a=0$ or $a=-5b$ we obtain case {\bf B} augmented with extra vector-like states
with opposite $U(1)_F$ charges.
Also, condition $\bar{b}\neq 0$ for case {\bf D} guarantees that we will not deal with case obtained from embedding of $SU(5)\tm U(1)_F$ in $SU(6)$ group.
Indeed, with $\bar{b}=0$ together with states $10_{-3\bar{a}}+10_{3\bar{a}}+2\tm 1'_{0}$ (which do not contribute in $SU(5)^2\cdot U(1)_F$, $\l U(1)_F\r^3$
and ${\rm Tr}Q$ anomalies) we get set $10_{2\bar{a}}+\bar 5_{-4\bar{a}}+2\tm (\bar 5_{-\bar{a}}+1_{5\bar{a}})$. The latter field content can be obtained via $SU(6)$ embedding as follows.
Consider $SU(6)$  field content $15+2\tm \bar 6$ which is anomaly free. Decomposition of $15$ and $\bar 6$ under $SU(6)\to SU(5)\tm U(1)_6$ is
$15=10_2+5_{-4}$ and $\bar 6=\bar 5_{-1}+1_5$, where for $U(1)_6$ charges the normalization factor $1/\sqrt{60}$ is neglected. Now making replacement
$5_{-4}\to \bar 5_{-4}$ and adding the pair $10_{-3}+10_3$ the field content will remain anomaly free. Adding to these two neutral singlets ($1'_0$ taken two times)
we will get the field content of {\bf D} with $\bar{b}=0$.
Note that discussing embedding of $SU(5)\tm U(1)_F$ in unitary groups, we skipped the $SU(6)$ group. The reason was that the case {\bf C}, obtained
from $E_6$ embedding, includes the case of $SU(6)$ embedding. This is not surprising since one of $E_6$'s maximal subgroup is $SU(6)\tm SU(2)$ and
$27$ (of $E_6$) decomposition  $E_6\to SU(6)\tm SU(2)$ is $27=(15,1)+(\bar 6, 2)$. Taking in case {\bf C} $a=5/4$, $b=3/4$ we will get states
$10_2+5_{-4}+2\tm (\bar 5_{-1}+1_5)$. These are  obtained by $SU(6)\to SU(5)\tm U(1)_6$ decomposition of $15+2\tm \bar 6$. That's why consideration of
unitary groups has been started from $SU(7)$.

Before closing this section, let us mention that for case {\bf D}, in  constructing the $\bar Q_{sup}$ charges,  besides $Q_{U(1)_6}$ and $Q_{U(1)_7}$, one can also
use another $U(1)$s - either charge of $U(1)'$ or $U(1)''$, or both together. However, one should make sure that superposition is such that all anomalies
are zero. For example, use $U(1)''$  symmetry. Then instead of Eq. (\ref{superposition1}) we will have
$\bar Q_{sup}=\bar aQ_{U(1)_6}+\bar bQ_{U(1)_7}+\bar cQ_{U(1)''}$. To do this, we should pick up from set {\bf D}  $10+\bar 5+1 +\bar 5+\bar 5+1$ and
(according to last equation in (\ref{flipSO10})) assign $U(1)''$ charges $1, 1, 1, -2, -2, 4$ respectively to these states.
Remaining two $10$-plets can have $U(1)''$
charges $p$ and $-p$, while $U(1)''$ charges of remaining two singlets are $k$ and $-k$. Thus, the set with  (one simple possible) $\bar Q_{sup}$ charge assignment will look:
$$
10_{-3\bar{a}+3\bar{b}+p\bar{c}}+10_{3\bar{a}+3\bar{b}-p\bar{c}}+10_{2\bar{a}-4\bar{b}+\bar{c}}+\bar 5_{-4\bar{a}-4\bar{b}-2\bar{c}}+
 \bar 5_{-\bar{a}-\bar{b}+\bar{c}}+\bar 5_{-\bar{a}-\bar{b}-2\bar{c}}
 $$
 \beq
 +1_{5\bar{a}-\bar{b}+\bar{c}}+1_{5\bar{a}-\bar{b}+4\bar{c}}+1'_{6\bar{b}+k\bar{c}}+1'_{6\bar{b}-k\bar{c}}~,
 ~~~~~{\rm with}~~~~30\bar{a}(3+2p)=\bar{c}(2k^2+10p^2-27)~.
\la{D-prime}
\eeq
Relations between $\bar{a}, \bar{c}, k$ and $p$ (imposed for  $\bar{c}\neq 0$) given in (\ref{D-prime}) insures that all anomalies vanish. Clearly, with rational selection of $\bar{a}, k$ and $p$ the value of $\bar{c}$
also will be rational. The set given in Eq. (\ref{D-prime}) is one simple selection among several options and opens up many possibilities for
model building with realistic phenomenology.

\section{$SU(5)\tm U(1)_F$ Models}\label{sect:models}

For model building with $U(1)_F$ symmetry we list and discuss requirements which should be satisfied in order
to obtain phenomenologically viable and economical setups.

{\bf (a)} In total, we should have three $10$-plets of $SU(5)$, four $\bar 5$-plets and one $5$-plet.
Out of these multiplets three pairs of $(10 +\bar 5)$ are matter superfields (containing quark and lepton superfields, as given in Eq. (\ref{su5matter})). The
$5$-plet and one remaining $\bar 5$-plet are scalar superfields\footnote{The scalar superfield $\Si (24)$ (neutral under $U(1)_F$),
needed for the symmetry breaking $SU(5)\to G_{321}$, is also assumed.} which will be denoted by $H$ and $\bar H$ respectively.

 {\bf (b)} In order to have top quark Yukawa coupling $\lam_t\sim 1$, the $U(1)_F$ symmetry should allow coupling $10_310_3H$ at renormalizable level.
At the same time, all $10$-plets should have different $U(1)_F$ charges in order to
generate adequately suppressed hierarchies of $\lam_u/\lam_t$ and  $\lam_c/\lam_t$.

Since for $U(1)_F$ charge assignments we have options given in (\ref{A})-(\ref{D}), for building three generation models with $U(1)_F$ flavor symmetry
we can consider different combinations of these assignments. For
example, one pair of $10, \bar 5$-plets can have $U(1)_F$ assignment {\bf A} (of Eq. (\ref{A})), another pair of $10, \bar 5$-plets
can have assignment {\bf B} and third pair of $10, \bar 5$-plets can come from selection {\bf C}. This collection can be refereed as {\bf ABC} model.
This model involves three $10$-plets, four $\bar 5$-plets and one $5$-plet (satisfying requirement {\bf (a)}). Other collections, such as {\bf ABB},
{\bf BBB}, etc., are also possible. However, selections like {\bf ACC}, {\bf CCC}, etc. are not allowed since they would involve extra $5$-plet(s) (not satisfying
 requirement {\bf (a)}). Note, considering, say, {\bf ABB} model, for two sets of $10, \bar 5$ coming with {\bf B} charge assignments
 should be taken $\al $ and $\al'\neq \al $ (for satisfying requirement {\bf (b)}). At the same time, for this selection extra pair of $5, \bar 5$-plets should be introduced with opposite $U(1)_F$ charges.

 {\bf (c)} Upon model building, one should make sure that only one $U(1)$ (identified with $U(1)_F$) emerges. For instance, if {\bf ABC}
 model is considered, the parameters $\al, a, b$ should not be independent. They should be fixed as $\al =\fr{m_1}{n_1}\bt $,
 $a =\fr{m_2}{n_2}\bt $, $b=\fr{m_3}{n_3}\bt $ ($m_i, n_i$ are integers). This would avoid extra global $U(1)$ symmetries.

Summarizing, satisfying all this requirements, we will group models in following five classes:
\beqr
{\bf ABB} \hs{2cm}& {\bf BBB}\hs{2cm} &{\bf D} \nonumber \\
{\bf ABC} \hs{2cm} & {\bf BBC} \hs{2cm}&
\la{classes}
\eeqr
Each of these includes several possibilities. Clarification of varieties of these possibilities is in order.

\vs{0.2cm}
\hs{-0.4cm}$\bullet $ {\bf Model ABB}

In this case we combine sets given by Eqs. (\ref{A}) and  (\ref{B}), and take: $10_0+\bar 5_0$, $10_{\al }+\bar 5_{-3\al }+1_{5\al }$ and
$10_{\al'}+\bar 5_{-3\al'}+1_{5\al'}$.  In addition, we introduce the pair $5_{q}+\bar 5_{-q}$. Thus, for this class, the complete field content is:
\beqr
10_0+\bar 5_0~,&~~~~10_{\al }+\bar 5_{-3\al }+1_{5\al } \nonumber \\
10_{\al'}+\bar 5_{-3\al'}+1_{5\al'}~,&\hs{-0.5cm}5_{q}+\bar 5_{-q}~.
\la{ABBclass}
\eeqr
This selection is not unique. We can exchange  $5$-plet's  $U(1)_F$ charge with one of the $\bar 5$-plets' charge. With this,
anomaly cancellation conditions are not changed. Thus, for $U(1)_F$ charge of the $5$-plet, identified with Higgs superfield $H(5)$, we have three (qualitatively different) options $Q_H= 0, -3\al $ or $q$. In counting these options, we took into account that
the charge selection $Q_H=-3\al '$ does not differ from selection $Q_H=-3\al $ (former is obtained from the latter by substitution $\al \to \al'$).
 Also, the case with $Q_H=-q$ is obtained from case $Q_H=q$ by substitution $q\to -q$. From the (remaining) four $\bar 5$-plets one should be identified with the Higgs superfield $\bar H(\bar 5)$. For each given $Q_H$, one should count how many qualitatively different charge assignments is possible for $\bar H$.
One can make sure that  for the pair $(Q_H, Q_{\bar H})$ eight different possibilities  are allowed:
 \beq
 (Q_H, Q_{\bar H})^{(i)}=\left \{ (q, -q), (q, 0), (q, -3\al ), (0, q), (0, -3\al ), (-3\al, q), (-3\al , 0), (-3\al , -3\al')\right \}~,
 \la{ABB-QHHbar}
 \eeq
 where $i=1, 2, \cdots , 8$ numerates (indicates) the options for the charge assignment for $H$ and $\bar H$. Thus, the content {\bf ABB} of Eq. (\ref{ABBclass})
 forms class with these different charge assignments. To make clear which particular $U(1)_F$ charge assignment for $H, \bar H$ is considered, it is instructive to use notation {\bf ABB}$^{(i)}$. For instance, {\bf ABB}$^{(i=3)}$  would mean that we are taking $(Q_H, Q_{\bar H})^{(i=3)}=(q,-3\al )$
 (see Eq. (\ref{ABB-QHHbar})).

 \vs{0.2cm}
\hs{-0.4cm}$\bullet $ {\bf Model ABC}

In this case, we collect together sets of Eqs. (\ref{A}),  (\ref{B}) and  (\ref{C}). Thus, the field content is:
$$
10_0+\bar{5}_0~, ~~~ 10_{\al }+\bar{5}_{-3\al }+1_{5\al }~,
$$
\beq
10_{a+b}+\bar{5}_{a-3b}+1_{a+5b}+5_{-2a-2b}+\bar{5}'_{-2a+2b}+1'_{4a}~.
\la{ABCclass}
\eeq
Since (\ref{ABCclass}) includes three $10$-plets, four $\bar 5$'s and one $5$-plet, we do not need to introduce any additional
vector-like states. Also in this case, we can exchange $U(1)_F$ charge of $5$-plet with one of the $\bar 5$'s charge. It turns out that
here we will have the following $20$ possibilities for $(Q_H, Q_{\bar H})$ pair selection:
\beqr
\hs{-0.5cm}(Q_H, Q_{\bar H})^{(i)}&\hs{-0.5cm}=&\hs{-0.4cm} \left \{(-2a-2b, 2b-2a), (-2a-2b, 0), (-2a-2b, -3\al ), (-2a-2b, a-3b), \right.  \nonumber \\
&& \hs{-0.2cm}(0, -3\al ), (0, a-3b), (0, -2a-2b), (0, 2b-2a), (-3\al, 0), (-3\al, a-3b), \nonumber \\
&& \hs{-0.2cm}(-3\al, -2a-2b),(-3\al, 2b-2a),(a-3b,0),(a-3b,-3\al),(a-3b,-2a\hs{-0.7mm}-\hs{-0.8mm}2b),  \nonumber \\
&&\left.  \hs{-0.2cm} (a\hs{-0.7mm}-\hs{-0.8mm}3b,2b\hs{-0.7mm}-\hs{-0.8mm}2a),(2b\hs{-0.7mm}-\hs{-0.8mm}2a,0),(2b\hs{-0.7mm}-\hs{-0.8mm}2a,-3\al),
(2b\hs{-0.7mm}-\hs{-0.8mm}2a,a\hs{-0.7mm}-\hs{-0.8mm}3b),(2b\hs{-0.7mm}-\hs{-0.8mm}2a,-2a\hs{-0.7mm}-\hs{-0.8mm}2b)\right\} .
\la{ABC-QHHbar}
 \eeqr
Thus, this {\bf ABC}$^{(i)}$ ($i=1,2,\cdots ,20$) class unifies twenty possible charge assignments for the pair $(H,~\bar H)$.

\vs{0.2cm}
\hs{-0.4cm}$\bullet $ {\bf Model BBB}

For constructing this case, we pick up the set of Eq. (\ref{B}) three times (with corresponding charge assignments) and
add the pair $5_{q}+\bar 5_{-q}$. Thus, the complete content is:
$$
10_{\al }+\bar{5}_{-3\al }+1_{5\al }~,~~~~~~10_{\al'}+\bar{5}_{-3\al' }+1_{5\al'}~,
$$
\beq
\hs{-1.8cm}10_{\al''}\hs{-1mm}+\hs{-1mm}\bar{5}_{-3\al''}\hs{-1mm}+\hs{-1mm}1_{5\al''}~,\hs{0.8cm}5_{q}+\bar 5_{-q}~.
\la{BBBclass}
\eeq
Here for $(Q_H, Q_{\bar H})$ pair selection we have four qualitatively different cases:
\beq
(Q_H, Q_{\bar H})^{(i)}=\left \{(q, -q), (q, -3\al), (-3\al, -3\al'),(-3\al, q) \right\}~.
\la{BBB-QHHbar}
\eeq
Therefore, this {\bf BBB}$^{(i)}$ ($i=1,\cdots ,4$) class unifies four options for the pair $(Q_H, Q_{\bar H})$.

\vs{0.2cm}
\hs{-0.4cm}$\bullet $ {\bf Model BBC}

The content for this case is build by taking set of (\ref{B}) two times with {\bf B}-type charge assignments, in combination of set
(\ref{C}). This gives the field content:
$$
\hs{-0.9cm}10_{\al }+\bar{5}_{-3\al }+1_{5\al }~, \hs{0.8cm} 10_{\al'}+\bar{5}_{-3\al'}+1_{5\al'}~,
$$
\beq
10_{a+b}+\bar{5}_{a-3b}+1_{a+5b}+5_{-2a-2b}+\bar{5}'_{-2a+2b}+1'_{4a}~.
\la{BBCclass}
\eeq
The list of possible $(Q_H, Q_{\bar H})$ pairs is:
\beqr
\hs{-0.5cm}(Q_H, Q_{\bar H})^{(i)}&\hs{-0.5cm}=&\hs{-0.4cm} \left \{(-2a-2b,-3\al), (-2a-2b, a-3b), (-2a-2b, 2b-2a), (2b-2a, -3\al), \right.  \nonumber \\
&& \hs{-0.2cm}(2b\hs{-0.7mm}-\hs{-0.8mm}2a, a\hs{-0.7mm}-\hs{-0.8mm}3b), (2b\hs{-0.7mm}-\hs{-0.8mm}2a, -2a\hs{-0.7mm}-\hs{-0.8mm}2b),
 (-3\al, -3\al'), (-3\al, a\hs{-0.7mm}-\hs{-0.8mm}3b), (-3\al, -2a\hs{-0.7mm}-\hs{-0.8mm}2b),  \nonumber \\
&&\hs{-0.2cm} \left. (-3\al, 2b-2a), (a-3b, -3\al),(a-3b,-2a-2b),(a-3b,2b-2a) \right\},
\la{BBC-QHHbar}
 \eeqr
giving thirteen possibilities unified in this {\bf BBC}$^{(i)}$ ($i=1,2,\cdots ,13$) class.

\vs{0.2cm}
\hs{-0.4cm}$\bullet $ {\bf Model D}

The field content of this model is given in (\ref{D}). It includes three $10$ and three $\bar 5$-plets. So, we do not need to combine
this content with other ones, but must add to it the pair $5_q+\bar 5_{-q}$. If the $U(1)_F$ charge assignments are just those given in
(\ref{D}), then for the pairs $(Q_H, Q_{\bar H})$ we will have eight options. However, as already discussed, it is possible to build charge assignments
utilizing additional $U(1)$-charges, as was done in the example given in Eq. (\ref{D-prime}). The latter case offers 13 distinct options for
the pairs $(Q_H, Q_{\bar H})$. These, open up varieties for the model building. One example from this {\bf D}-class of models is presented in Sect. \ref{sect:D-prime}.

\subsection{Up-type Quark Yukawa Matrices}\label{sect:UpQuarkMs}

In order to proceed with model building, first we give all acceptable up-type Yukawa textures obtained by $U(1)_F$ symmetry. In our approach, among up-type
quarks only top quark has renormalizable Yukawa coupling. Yukawa couplings $\lam_u$ and $\lam_c$ emerge after $U(1)_F$ flavor symmetry breaking.
The breaking of $U(1)_F$ should be achieved by flavon superfields. Here we consider simple set of flavon pair $X+\bar X$ with $U(1)_F$ charges
\beq
Q(X)=-\bt ~,~~~~~~~~Q(\bar X)=\bt ~.
\la{XbarX-ch}
\eeq
 In general, scalar components of $X$ and $\bar X$ have different VEVs $\lan X\ran $ and $\lan \bar X\ran $ respectively.
Detailed discussion of possibility for $U(1)_F$ symmetry breaking, giving fixed VEVs for $X$ and $\bar X$, is presented in Appendix \ref{apA}. We introduce the
notations
\beq
\fr{|X|}{M_{\rm Pl}}= \ep ~,~~~~\fr{|\bar X|}{M_{\rm Pl}}= \bar{\ep }~,
\la{eps-epsbar}
\eeq
where $M_{\rm Pl}\simeq 2.4\cdot 10^{18}$~GeV is reduced Planck scale, which will be treated as natural cut off for all higher-dimensional
non-renormalizable operators.
Thus, the hierarchies between Yukawa couplings and CKM mixing angles will be expressed by powers of small parameters
$\ep, \bar{\ep }\ll 1$.

Due to the composition of the $10$-plet given in Eq. (\ref{su5matter}) and taking into account that $H(5)\supset h_u$, the up-type quark
masses emerge through the Yukawa couplings of the form $10\cdot 10\cdot H$, where family and $SU(5)$ indices are suppressed.
As it turns out, within this setup, three acceptable Yukawa textures emerge for up-type quarks. These textures will be referred as {\bf U1},
{\bf U2} and {\bf U3}.

\vs{0.2cm}
{\bf (i) Up Quark Yukawa Texture U1}
\vs{0.2cm}

The $U(1)_F$ charges of three $10$-plets and the Higgs superfield $H$ are:
\beq
Q(10_1)=n\bt \l n\bt+\bt \r  ~,~~Q(10_2)=n\bt -\bt ~,~~Q(10_3)=n\bt -3\bt ~,~~Q(H)=6\bt -2n\bt ~.
\la{QU1}
\eeq
This selection provides the following Yukawa texture
\beq
\begin{array}{ccc}
 & {\begin{array}{ccc}
\hs{-0.4cm}10_1 &~~~ 10_2  ~~~&10_3 \hs{0.2cm}
\end{array}}\\ \vspace{1mm}
{\bf U1}:~~~
\begin{array}{c}
 10_1\\ 10_2 ~  \\ 10_3 ~
 \end{array}\!\!\!\!\!\hs{-0.1cm}&{\!\left(\begin{array}{ccc}

 \ep^6(\ep^8)&\ep^5(\ep^6)&\ep^3(\ep^4)
\\
 \ep^5(\ep^6)&\ep^4 &\ep^2
 \\
\ep^3(\ep^4)&\ep^2  &1
\end{array}\right)H}~,
\end{array}  \!\!  ~~~
\label{U1}
\eeq
where dimensionless couplings (whose magnitudes are assumed to be $\sim 1/3 -3$) are not displayed. With $\ep=1/10-1/5$, the matrix (\ref{U1})
gives right hierarchies between up-type quark Yukawas.

\vs{0.2cm}
{\bf (ii) Up Quark Yukawa Texture U2}
\vs{0.2cm}

In this case we use the following assignment
\beq
Q(10_1)=n\bt +3\bt ~,~~Q(10_2)=n\bt  ~,~~Q(10_3)=n\bt +\bt ~,~~Q(H)=-2n\bt -2\bt ~,
\la{QU2}
\eeq
which gives the texture:
\beq
\begin{array}{ccc}
 & {\begin{array}{ccc}
\hs{-0.4cm}10_1 & 10_2  &10_3 \hs{0.2cm}
\end{array}}\\ \vspace{1mm}
{\bf U2}:~~~
\begin{array}{c}
 10_1\\ 10_2 ~  \\ 10_3 ~
 \end{array}\!\!\!\!\!\hs{-0.1cm}&{\!\left(\begin{array}{ccc}

 \ep^4& ~~\ep  & ~~\ep^2
\\
 \ep &~~ \bar{\ep }^{\hs{0.1cm}2} & ~~ \bar{\ep }
 \\
\ep^2& ~~ \bar{\ep }  &~~ 1
\end{array}\right)H}~.
\end{array}  \!\!  ~~~
\label{U2}
\eeq
With selection $\bar{\ep }=1/10-1/20$, $\ep \sim (1/5-1/10)\cdot \bar{\ep}^2$, the needed hierarchies
for the ratios $\lam_u/\lam_c$, $\lam_c/\lam_t$ are generated.

\vs{0.2cm}
{\bf (iii) Up Quark Yukawa Texture U3}
\vs{0.2cm}

Finally, with $U(1)_F$ charge selections
\beq
Q(10_1)=n\bt -\bt ~,~~Q(10_2)=n\bt  ~,~~Q(10_3)=n\bt +\bt ~,~~Q(H)=-2n\bt -2\bt ~,
\la{QU3}
\eeq
the up-type quark Yukawa couplings will be
\beq
\begin{array}{ccc}
 & {\begin{array}{ccc}
\hs{-0.5cm}10_1 & 10_2  &10_3 \hs{0.2cm}
\end{array}}\\ \vspace{1mm}
{\bf U3}:~~~
\begin{array}{c}
 10_1\\ 10_2 ~  \\ 10_3 ~
 \end{array}\!\!\!\!\!\hs{-0.1cm}&{\!\left(\begin{array}{ccc}

 \bar{\ep }^{\hs{0.1cm}4}& ~~ \bar{\ep }^{\hs{0.1cm}3}  & ~~ \bar{\ep }^{\hs{0.1cm}2}
\\
 \bar{\ep }^{\hs{0.1cm}3} &~~ \bar{\ep }^{\hs{0.1cm}2} & ~~ \bar{\ep }
 \\
\bar{\ep }^{\hs{0.1cm}2} & ~~ \bar{\ep }  &~~ 1
\end{array}\right)H}~,
\end{array}  \!\!  ~~~
\label{U3}
\eeq
which for $\bar{\ep}\sim 1/20-1/10$ gives successful explanation of  hierarchies
$\lam_u/\lam_c\sim \bar{\ep}^2$ and $\lam_c/\lam_t\sim \bar{\ep}^2$.

This classification of up-type Yukawa textures helps to build models emerging from classes of Eq. (\ref{classes})
(for each class, see discussion after Eq. (\ref{classes})). As one can see, there are many possibilities to be considered in order
to see which one gives phenomenologically viable model. Detailed investigation and complete list of acceptable scenarios will be presented
in a longer paper \cite{inprep}. Below we present three models with successful explanation of hierarchies between charged fermion
masses and mixings.

\subsection{{\bf ABC$^{(i=4)}$-U1$_{(n=1)}$} Model}\la{abc}

In this model, the content of Eq. (\ref{ABCclass}) is considered and  charges are matched in such a way as to obtain with up-type Yukawa texture {\bf U1}
of Eq. (\ref{U1}).
Here, selection $n=1$ is made. Thus, according to Eq. (\ref{QU1}), the charge of $H(5)$-plet is $Q_H=4\bt$, while charges of
$10$-plets are $Q(10_i)=\{\bt, 0, -2\bt \}$. From the set (\ref{ABCclass}) we will identify $10_{\al }$, $10_0$ and $10_{a+b}$ with $1^{\rm st}$,
$2^{\rm nd}$ and $3^{\rm rd}$ families respectively, and $5_{-2a-2b}$ with $H$. Making the charge matching
$\al =\bt $, $a+b=-2\bt $ and selection $a=-\bt/4$, we will have
\beq
\{\al, ~a, ~b\}=\{\bt, ~-\bt/4,~ -7\bt/4\}~.
\la{select-abal}
\eeq
Furthermore, since we are dealing with {\bf ABC$^{(i=4)}$} model, using Eqs.  (\ref{ABC-QHHbar}), (\ref{select-abal}) we have
$(Q_H,Q_{\bar H})^{(i=4)}$ $=(-2a-2b, a-3b)=(4\bt, 5\bt)$. The charges of remaining $\bar 5$-plets: $\bar 5_0$, $\bar 5_{-3\al}$
and $\bar 5_{-2a+2b}$, which we identify with $1^{\rm st}$, $2^{\rm nd}$ and $3^{\rm rd}$ families of matter $\bar 5$-plets respectively,
will be $Q(\bar 5_i)=\{0, -3\bt, -3\bt \}$. The $U(1)_F$ charge assignment of all states of content (\ref{ABCclass}) is summarized in Table \ref{t:tab1}.
%
%
\begin{table}
\caption{$U(1)_F$ charge assignment for {\bf ABC$^{(i=4)}$-U1$_{(n=1)}$} model.
 }

\label{t:tab1} $$\begin{array}{|c||c|c|c|c|c|c|c|c|c|c|c|}

\hline
\vs{-0.3cm}
 &  &  &  &  &  &  &  & & &&\\

\vs{-0.4cm}

& ~10_1~& ~10_2~&~ 10_3~& ~\bar 5_1 ~& ~\bar 5_2~ &~\bar 5_3~&~H(5) ~&~ \bar H(\bar 5) ~  & ~1_1 ~ & ~1_2 & 1_3\\

&  &  &  &  &  &  &  &  & &&\\

\hline

\vs{-0.3cm}
 &  &  &  &  &  &  &  & & &&\\

\vs{-0.3cm}
~Q_{U(1)_F}~& \bt &0  &-2\bt  & 0  & -3\bt   &-3\bt   & 4\bt   & 5\bt  & -\bt   &5\bt &-9\bt \\

&  &  &  &  &  &  &  & & &&\\

\hline

\end{array}$$

\end{table}
%
%
%
%
%
With this assignment, the Yukawa coupling matrices are determined as follows:
\beq
\begin{array}{ccc}
 & {\begin{array}{ccc}
\hs{-0.8cm}10_1 &10_2  &~10_3 \hs{0.2cm}
\end{array}}\\ \vspace{1mm}
\begin{array}{c}
 10_1\\ 10_2 ~  \\ 10_3 ~
 \end{array}\!\!\!\!\!\hs{-0.1cm}&{\!\left(\begin{array}{ccc}

 \ep^6& ~~\ep^5  & ~~\ep^3
\\
 \ep^5 &~~ \ep ^{\hs{0.1cm}4} & ~~ \ep ^{\hs{0.1cm}2}
 \\
\ep^3& ~~ \ep^2  &~~ 1
\end{array}\right)H}~,~~~
\end{array}  \!\!  ~~~
\begin{array}{ccc}
 & {\begin{array}{ccc}
\hs{-0.6cm}\bar 5_1 &~~ \bar 5_2  &~~\bar 5_3 \hs{0.2cm}
\end{array}}\\ \vspace{1mm}
\begin{array}{c}
 10_1\\ 10_2 ~  \\ 10_3 ~
 \end{array}\!\!\!\!\!\hs{-0.1cm}&{\!\left(\begin{array}{ccc}

 \ep^6& ~~\ep^3  & ~~\ep^3
\\
 \ep^5 &~~ \ep ^{\hs{0.1cm}2} & ~~ \ep ^{\hs{0.1cm}2}
 \\
\ep^3& ~~ 1  &~~ 1
\end{array}\right)\bar H}~.
\end{array}  \!\!  ~~~
\label{UDE-abc-u1}
\eeq
Taking into account Eq. (\ref{su5matter}) and $H\supset h_u, ~\bar H\supset h_d$, Eq. (\ref{UDE-abc-u1}) yield:
$$
\lam_u:\lam_c:\lam_t\sim \ep^6:\ep^4:1~,~~~~\lam_t\sim 1~,
$$
\beq
\lam_e:\lam_{\mu }:\lam_{\tau }\sim \ep^6:\ep^2:1~, ~~~~~\lam_d:\lam_s:\lam_b\sim \ep^6:\ep^2:1~.
\la{UDE-hier-abc}
\eeq
Assuming that in (\ref{UDE-abc-u1}) there are dimensionless Yukawa couplings with natural values - in a range $\sim 1/3-3$, with selection $\ep \simeq 0.2$,
the hierarchies in (\ref{UDE-hier-abc}) can fit well with the experimental data. Notice that $\lam_{b, \tau }\sim 1$, which means that
in this scenario $\tan \bt \approx 55-60$.
As far as the CKM mixing angles are concerned, from (\ref{UDE-abc-u1}) one can obtain:
\beq
|V_{us}|\sim \ep ~,~~~|V_{cb}|\sim \ep^2 ~,~~~|V_{ub}|\sim \ep^3~.
\la{CKM-abc}
\eeq
These are also of right magnitudes (with $\ep \simeq 0.2$).
Because of the charge equality $Q(\bar 5_2)=Q(\bar 5_3)$, corresponding entries in $2^{\rm nd}$ and $3^{\rm rd}$ columns of the second matrix of Eq. (\ref{UDE-abc-u1}) have comparable sizes. Taking into account that $\bar 5\supset l$, this leads to the naturally large mixing between $l_2$ and $l_3$ lepton flavors:
\beq
\tan \te_{\mu \tau }\sim 1~,
\la{mu-tau-abc}
\eeq
providing good explanation for large $\nu_{\mu }-\nu_{\tau }$ neutrino oscillations.
To demonstrate this, we also discuss neutrino sector in some extent.
Let us  work in a basis where the matrix responsible for the charged lepton masses ($2^{\rm nd}$ matrix in Eq. (\ref{UDE-abc-u1})) is diagonal. Thus, the mixing matrix emerging from the neutrino sector will coincide with lepton mixing matrix. We will apply the singlet state $1_1$ (see Tab. \ref{t:tab1}) as a right-handed
neutrino. The relevant couplings are
$\lam_{\nu}(\bar 5_3+t\bar 5_2)1_1H+\hat{M}\ov{\ep}^21_11_1$, with $\lam_{\nu}, t$ being dimensionless couplings and $\hat{M}$ some scale.
Moreover, we also include higher order operators $\lam_1\ep^5\bar 5_1\bar 5_2HH/M'$ and $\lam_2\ep^2\bar 5_2\bar 5_2HH/M''$.
Integration out of the state $1_1$, together with latter operators, give the neutrino mass matrix:
\vs{-0.5cm}
\beq
\begin{array}{ccc}
 & {\begin{array}{ccc}
 &  &
\end{array}}\\ \vspace{1mm}
\begin{array}{c}
 \\  ~  \\ ~
 \end{array}\!\!\!\!\!\hs{-0.1cm}&{\!M_{\nu}=\left(\begin{array}{ccc}

 0& ~~0  & ~~0
\\
 0 &~~ t^2 & ~~ t
 \\
0& ~~ t  &~~ 1
\end{array}\right)}m~+
\end{array}  \!\!
\begin{array}{ccc}
 & {\begin{array}{ccc}
\hs{-0.6cm} &~~   &~~ \hs{0.2cm}
\end{array}}\\ \vspace{1mm}
\begin{array}{c}
 \\  ~  \\  ~
 \end{array}\!\!\!\!\!\hs{-0.1cm}&{\!\hs{-0.5cm}\left(\begin{array}{ccc}

 0& ~~1  & ~~0
\\
 1 &~~ \de & ~~ 0
 \\
0& ~~ 0  &~~0
\end{array}\right)\!\ov{m}}~,
\end{array}  \!\!  ~~~
\label{Mnu-abc-u1}
\eeq
with $m=\fr{\lam_{\nu}^2v_u^2}{\hat{M}\ov{\ep}^2}$, $\ov{m}=\fr{\lam_1\ep^5}{M'}v_u^2$ and $\de =\fr{\lam_2M'}{\lam_1M''\ep^3}$.
The first matrix at r.h.s. of (\ref{Mnu-abc-u1}) (emerged by integrating out the $1_1$ state) is mostly responsible for
the mass $m_{\nu_3}$ and leptonic $\te_{23}$ mixing. Indeed, in the limit $\ov{m}\to 0$, we get $\tan \te_{23}=|t|$. This, for $|t|\sim 1$ (natural value), gives
$\te_{23}\approx 45^o$. Inclusion of the $\ov{m}$ terms are responsible for mixing angles $\te_{12}, \te_{13}$ and masses $m_{\nu_{1,2}}$.
With a selection $m=0.029$~eV, $\ov{m}=0.0116$~eV, $t=0.78$, $\de=0.8$ we obtain $\De m_{\rm atm}^2=m_{\nu_3}^2-m_{\nu_2}^2\simeq 2.6\cdot 10^{-3}~{\rm eV}^2$,
$\De m_{\rm sol}^2=m_{\nu_2}^2-m_{\nu_1}^2\simeq 7.2\cdot 10^{-5}~{\rm eV}^2$, $\te_{12}=34^o$, $\te_{23}=45.3^o$, $\te_{13}=9^o$.
These agree well with a recent data \cite{Schwetz:2011zk}. In this considered case neutrinos are hierarchical in mass:
$m_{\nu_i}=(0.00688,~0.01093,~0.052)$~eV. The values of parameters used above are obtained with $\lam_{\nu,1,2}\sim 1$, $\ov{\ep}\simeq 0.25$,
$\hat{M}\sim 10^{16}$~GeV, $M'\sim 10^{12}$~GeV,  $M''\sim 10^{14}$~GeV. Although the values of these scales remain unexplained within this scenario, we have showed that the model can be compatible with neutrino sector. More detailed study of this and related issues will be presented in \cite{inprep}.

As in minimal $SU(5)$ GUT, some care is needed to cure the problem of $M_D-M_E$ mass degeneracy. For fixing this problem one can use either an extension
by scalar $45$-supermultiplets \cite{Georgi:1979df}, or include powers of adjoint $24$-plet in the Yukawa couplings \cite{Ellis:1979fg},
or utilize extra heavy matter supermultiplets \cite{Shafi:1999rm}. Study of this problem is beyond the scope of this Letter.

Before closing this subsection, let us mention that within this scenario the splitting between masses of doublets and triplets
(coming from $H, \bar H$) should be obtained via fine tuning
(as in minimal SUSY $SU(5)$) of the model parameters. However, one should make sure that this is possible to achieve. Due to the $U(1)_F$ symmetry,
renormalizable superpotential couplings $(M_H+\lam_H\Si )H\bar H$ are forbidden. However, in this scenario we have extra $SU(5)$ singlet
states charged under $U(1)_F$ (see Table \ref{t:tab1}). For instance, picking up the states $1_2$ and $1_3$ and announcing them as scalar superfields
(with positive matter $R$-parity), the relevant lowest superpotential couplings (including them) will be
$M_{\rm Pl}^2\ep^51_2+M_{\rm Pl}^2\bar{\ep}^91_3+M_{\rm Pl}\bar{\ep}^41_21_3$, where dimensionless couplings have been neglected (assuming that they are
of the order of unity). One can check that vanishing of the $F$-terms $F_{1_2}=F_{1_3}=0$ lead to the induced VEVs
$\lan 1_2\ran \sim M_{\rm Pl}\bar{\ep}^5$ and $\lan 1_3\ran \sim M_{\rm Pl}\ep^5/\bar{\ep}^4$. With selection $\bar{\ep}\sim 0.25$ we will have
$\lan 1_2\ran \sim 10^{-3}M_{\rm Pl}$, $\lan 1_3\ran \sim 0.1M_{\rm Pl}$ without affecting anything in the discussion above.
However, the couplings $1_3(\lam_H+{\lam'}_H\fr{\Si}{M_{\rm Pl}})H\bar H$ with $\lan \Si \ran =V\cdot {\rm Diag}(2, 2, 2, -3,-3)$ and tuning
condition $\lam_H=3{\lam'}_HV/M_{\rm Pl}$ (satisfied with $\lam_H\sim 0.1$, ${\lam'}_H\sim 2-8$, rendering theory self consistent) lead to the
massless doublets ($M_{H_2}=0$) and colored triplets with masses $M_{H_3}=\fr{5}{3}\lam_H\lan 1_3\ran \sim {\rm few}\cdot M_{\rm GUT}$.

\subsection{{\bf BBB$^{(i=2)}$-U3$_{(n=-2/5)}$} Model}\la{bbb}


Within {\bf BBB} model with content (\ref{BBBclass}), one successful scenario is obtained with
$i=2$ (in Eq. (\ref{BBB-QHHbar})) and with up-type Yukawa texture {\bf U3} with $n=-2/5$ (see Eqs. (\ref{QU3}), (\ref{U3})). For
$10$-plets charges we will make the following matching $Q(10_i)=\{n\bt-\bt, n\bt, n\bt+\bt \}=\{\al'', \al', \al \}$. With this, using Eqs. (\ref{BBB-QHHbar}), (\ref{QU3}) we will have $(Q_H,Q_{\bar H})^{(i=2)}=(q, -3\al )=(-2n\bt-2\bt, -3n\bt-3\bt)$. Remaining $\bar 5$-plets, $\bar 5_{-3\al''}$, $\bar 5_{-3\al'}$
and $\bar 5_{-q}$ will be identified as $1^{\rm st}$, $2^{\rm nd}$ and  $3^{\rm rd}$ families respectively of the matter $\bar 5$ states. Therefore,
$Q(\bar 5_i)=\{3\bt-3n\bt, -3n\bt, 2\bt+2n\bt \}$. With selection $n=-2/5$, the $U(1)_F$ charges of all states from the content  (\ref{BBBclass}) are given in
Table \ref{t:tab2}.
%
%
\begin{table}
\caption{$U(1)_F$ charge assignment for {\bf BBB$^{(i=2)}$-U3$_{(n=-2/5)}$} model.
 }

\label{t:tab2} $$\begin{array}{|c||c|c|c|c|c|c|c|c|c|c|c|}

\hline
\vs{-0.3cm}
 &  &  &  &  &  &  &  & & &&\\

\vs{-0.4cm}

& ~10_1~& ~10_2~&~ 10_3~& ~\bar 5_1 ~& ~\bar 5_2~ &~\bar 5_3~&~H(5) ~&~ \bar H(\bar 5) ~  & ~1_1 ~ & ~1_2 & 1_3\\

&  &  &  &  &  &  &  &  & &&\\

\hline

\vs{-0.3cm}
 &  &  &  &  &  &  &  & & &&\\

\vs{-0.3cm}
~Q_{U(1)_F}~& -\fr{7}{5}\bt &-\fr{2}{5}\bt &\fr{3}{5}\bt &\fr{21}{5}\bt &\fr{6}{5}\bt &\fr{6}{5}\bt &-\fr{6}{5}\bt &-\fr{9}{5}\bt &-7\bt &-2\bt &3\bt \\

&  &  &  &  &  &  &  & & &&\\

\hline

\end{array}$$

\end{table}
%
%
%
With these assignments, couplings responsible for up-type quark Yukawas are given in Eq. (\ref{U3}), while couplings generating charged lepton and down
quark masses are:
\beq
\begin{array}{ccc}
 & {\begin{array}{ccc}
\hs{-0.6cm}\bar 5_1 &~~ \bar 5_2  &~~\bar 5_3 \hs{0.2cm}
\end{array}}\\ \vspace{1mm}
\begin{array}{c}
 10_1\\ 10_2 ~  \\ 10_3 ~
 \end{array}\!\!\!\!\!\hs{-0.1cm}&{\!\left(\begin{array}{ccc}

 \ep & ~~\bar{\ep }^2 & ~~\bar{\ep }^2
\\
 \ep^2 &~~ \bar{\ep } & ~~ \bar{\ep }
 \\
\ep^3& ~~ 1  &~~ 1
\end{array}\right)\bar H}~.
\end{array}  \!\!  ~~~
\label{DE-bbb-u3}
\eeq
These (with $\ep <\bar{\ep}$) give
$$
\lam_u:\lam_c:\lam_t\sim \bar{\ep }^4:\bar{\ep }^2:1~,~~~~\lam_t\sim 1~,
$$
\beq
\lam_e:\lam_{\mu }:\lam_{\tau }\sim \ep :\bar{\ep }:1~, ~~~~~\lam_d:\lam_s:\lam_b\sim \ep :\bar{\ep }:1~.
\la{UDE-hier-bbb}
\eeq
Taking $\bar{\ep }\sim 1/20-1/10$ and $\ep\sim 3\cdot 10^{-4}$, the pattern (\ref{UDE-hier-bbb}) describe well hierarchies between charged fermion
Yukawa couplings. Also, the CKM matrix elements are properly suppressed: $|V_{us}|\sim \bar{\ep }, |V_{ub}|\sim \bar{\ep }^2, |V_{cb}|\sim \bar{\ep }$.
Because of the large mixing between $\bar 5_2$ and $\bar 5_3$ states, also in this case for leptonic mixing we expect $\tan \te_{\mu \tau}\sim 1$,
providing large $\nu_{\mu }-\nu_{\tau }$ oscillations. Demonstration of this can be done in a same way as for the model presented in Sect.
\ref{abc}.

The doublet-triplet splitting within this scenario can be achieved in the same manner as was discussed at the end of the Sect. \ref{abc} for
{\bf ABC$^{(i=4)}$-U1$_{(n=1)}$} model. Without going in this discussion, let us proceed to consider
another scenario.

\subsection{{\bf D-U3$_{(n=1)}$} Model: Content of Eq. (\ref{D-prime})}\la{sect:D-prime}

The field content of this model is given in Eq. (\ref{D-prime}) augmented with Higgs superfields $H(5)$ and $\bar H(\bar 5)$ of $U(1)_F$
charges $q$ and $-q$ respectively. We will match charges of the $10$-plets with assignments of {\bf U3} texture (see Eq. (\ref{QU3})) as
follows $Q(10_i)=\{n\bt-\bt, n\bt, n\bt+\bt \}=\{2\bar{a}-4\bar{b}+\bar{c},  3\bar{a}+3\bar{b}-p\bar{c},  -3\bar{a}+3\bar{b}+p\bar{c}\}$.
Therefore, $q=-2n\bt-2\bt$. A phenomenologically viable model is obtained with the selection $p=k=-8/3$. This, with the matching given above and condition in
Eq. (\ref{D-prime}), give $(\bar{a}, \bar{b}, \bar{c}, n)=\l \fr{5}{2}\bt , \fr{1}{2}\bt , -3\bt , 1\r$. Furthermore, we make the identification of flavors of
$\bar 5$-plets as:
$(\bar 5_1, \bar 5_2, \bar 5_3)=(\bar 5_{-\bar{a}-\bar{b}-2\bar{c}}~, \bar 5_{-4\bar{a}-4\bar{b}-2\bar{c}}~, \bar 5_{-\bar{a}-\bar{b}+\bar{c}})$.
With these selections and parameters determined above, all $U(1)_F$ charges get fixed (in the unit of $\bt$). In Table \ref{t:tab3}
we summarize the charges of all states.\footnote{Since with this assignment $1_2$'s $U(1)_F$ charge is zero and it does not contribute to anomalies,
there is no need for introducing state $1_2$. However, presence of four singlets (including $1_2$) is required with more general assignment of Eq. (\ref{D-prime}).}
%
%
%
\begin{table}
\caption{$U(1)_F$ charge assignment for {\bf D-U3$_{(n=1)}$} model with content of Eq. (\ref{D-prime}).
For parameters the following selection is made $(\bar{a}, \bar{b}, \bar{c})=\l \fr{5}{2}\bt , \fr{1}{2}\bt , -3\bt\r$, $p=k=-8/3$.
 }

\label{t:tab3} $$\begin{array}{|c||c|c|c|c|c|c|c|c|c|c|c|c|}

\hline
\vs{-0.3cm}
 &  &  &  &  &  &  &  & & &&&\\

\vs{-0.4cm}

& ~10_1~& ~10_2~&~ 10_3~& ~\bar 5_1 ~& ~\bar 5_2~ &~\bar 5_3~&~H(5) ~&~ \bar H(\bar 5) ~  & ~1_1 ~ & ~1_2 & 1_3&1_4\\

&  &  &  &  &  &  &  &  & &&&\\

\hline

\vs{-0.3cm}
 &  &  &  &  &  &  &  & & &&&\\

\vs{-0.3cm}
~Q_{U(1)_F}~& 0 & \bt & 2\bt & 3\bt & -6\bt & -6\bt &-4\bt &4\bt &9\bt &0 &11\bt &-5\bt \\

&  &  &  &  &  &  &  & & &&&\\

\hline

\end{array}$$

\end{table}
%
%
%
With these assignments, the Yukawa couplings are:
\beq
\begin{array}{ccc}
 & {\begin{array}{ccc}
\hs{-0.9cm}10_1 &10_2  &~10_3 \hs{0.2cm}
\end{array}}\\ \vspace{1mm}
\begin{array}{c}
 10_1\\ 10_2 ~  \\ 10_3 ~
 \end{array}\!\!\!\!\!\hs{-0.1cm}&{\!\left(\begin{array}{ccc}

 \bar{\ep}^4& ~~\bar{\ep}^3  & ~~\bar{\ep}^2
\\
 \bar{\ep}^3 &~~ \bar{\ep}^{\hs{0.1cm}2} & ~~ \bar{\ep}
 \\
\bar{\ep}^2& ~~ \bar{\ep}  &~~ 1
\end{array}\right)H}~,~~~
\end{array}  \!\!  ~~~
\begin{array}{ccc}
 & {\begin{array}{ccc}
\hs{-0.6cm}\bar 5_1 &~~ \bar 5_2  &~~\bar 5_3 \hs{0.2cm}
\end{array}}\\ \vspace{1mm}
\begin{array}{c}
 10_1\\ 10_2 ~  \\ 10_3 ~
 \end{array}\!\!\!\!\!\hs{-0.1cm}&{\!\left(\begin{array}{ccc}

 \ep^7& ~~\bar{\ep}^2  & ~~\bar{\ep}^2
\\
 \ep^8 &~~ \bar{\ep} & ~~\bar{\ep}
 \\
\ep^9& ~~ 1  &~~ 1
\end{array}\right)\bar H}~.
\end{array}  \!\!  ~~~
\label{UDE-d-u1}
\eeq
These textures lead to:
$$
\lam_u:\lam_c:\lam_t\sim \bar{\ep }^4:\bar{\ep }^2:1~,~~~~\lam_t\sim 1~,
$$
\beq
\lam_e:\lam_{\mu }:\lam_{\tau }\sim \ep^7 :\bar{\ep }:1~, ~~~~~\lam_d:\lam_s:\lam_b\sim \ep^7 :\bar{\ep }:1~.
\la{UDE-hier-d}
\eeq
With $\bar{\ep }\sim 1/20-1/10$ and $\ep \approx 0.3$, the ratios in Eq. (\ref{UDE-hier-d}) describe well observed hierarchies between charged fermion masses.
Also, the CKM mixing angles have adequately suppressed values: $|V_{us}|\sim \bar{\ep }, |V_{ub}|\sim \bar{\ep }^2, |V_{cb}|\sim \bar{\ep }$, while for the leptonic mixing angle $\te_{\mu \tau}$ one
expects $\tan \te_{\mu \tau}\sim 1$ (as was demonstrated for the model presented in Sect. \ref{abc}).

Since in this scenario the superfields $H$ and $\bar H$ have opposite $U(1)_F$ charges, the doublet-triplet splitting can
be obtained in the same way (by fine tuning) as within minimal SUSY $SU(5)$. Thus, no additional effort is needed, unlike the scenarios considered
in Sections \ref{abc} and \ref{bbb}.

Finally, let us note that by proper shift of $U(1)_F$ charges of the states of Table \ref{t:tab3}, one can obtain the charge assignments
of model {\bf BBB$^{(i=2)}$-U3$_{(n=-2/5)}$} given in  Tab. \ref{t:tab2}. However, the latter's assignment leads to different
phenomenology (such as the different couplings required for the doublet-triplet splitting etc.). That's why, as a different model, this scenario
has been presented separately.

\vs{0.3cm}

Since within considered scenarios matter superfields ($f_i$) have family dependent $U(1)_F$ charges $Q_{f_i}$, there is potentially new source
for sfermion mass non-universality. In particular, as given at the end of Appendix, after SUSY breaking $D_{U(1)_F}$-term becomes
$2(m_X^2-m_{\bar X}^2)/\tl{g}^2$, where $\tl{g}$ is $U(1)_F$'s coupling constant and $m_X^2$ and $m_{\bar X}^2$ are soft mass$^2$'s of the
scalar components of the flavon superfields $X$ and $\bar X$ respectively. Non-zero $D_{U(1)_F}$-term give non-universal contribution to the sfermion
masses of the form $\De m_{\tl{f}_i}^2=Q_{f_i}(m_X^2-m_{\bar X}^2)/2$ raising new source for FCNC. However, within minimal $N=1$
SUGRA \cite{Chamseddine:1982jx}, due to
$m_X^2=m_{\bar X}^2$ universality, this contribution vanish and we have no additional source for flavor violation. Note that the relation $m_X^2=m_{\bar X}^2$
is quite stable against radiative corrections. Only couplings which might affect this relation, via loop corrections, are couplings
of $X$ and  $\bar X$ states with matter. However, this kind of couplings, appearing at high-dimensional operator level, are strongly suppressed. This insures stability of
the relation  $m_X^2=m_{\bar X}^2$. Note also that below the $U(1)_F$ symmetry breaking scale the $D_{U(1)_F}$-term is not renormalized.
Therefore, we conclude that in order to avoid new contributions to the FCNC (which is common problem within generic SUGRA) one should work within
framework (such as minimal SUGRA) giving universality of soft masses.

\section{Conclusions}\la{sec:conc}

In this Letter we have presented new examples of non-anomalous flavor $U(1)_F$ symmetries within SUSY $SU(5)$ GUT. Our way
of finding of such $U(1)_F$s was to embed the $SU(5)\tm U(1)_F$ in non-Abelian group with anomaly free content. Our selection was based on the
requirement that non-trivial $SU(5)$ states should be just those of minimal SUSY $SU(5)$, while the number of additional singlet states should
not be large. The latter, within concrete scenario, can be exploited for model building  with realistic phenomenology. For demonstrative purposes
we have presented three models which nicely explain hierarchies between charged fermion masses and mixings. We have not addressed the problem of
wrong asymptotic mass relations $M_D=M_E^T$, common also for minimal $SU(5)$ GUT. Solution of this problem can be achieved either by inclusion of scalar
$45$ supermultiplets \cite{Georgi:1979df}, or appropriate powers of the Higgs supermultiplet of $24$ (adjoint) in the Yukawa interactions \cite{Ellis:1979fg},
or specific extension of the matter sector \cite{Shafi:1999rm} can be considered. Within the models, we have found, many varieties
of possibilities emerge which require detailed investigation. Complete study of these,  together with neutrino sector (some of the singlets, involved in the
considered models, can serve as right-handed neutrinos) and other phenomenological issues will be presented in forthcoming publication \cite{inprep}.

\vspace*{0.3in}
\hs{-0.6cm}\textbf{Acknowledgement}

\vs{0.2cm}
\hs{-0.6cm}I thank K.S. Babu and M.C. Chen for useful comments, and interesting and stimulating discussions.
I wish to thank the Center for Theoretical Underground Physics and Related Areas (CETUP*) in Lead, South Dakota and
the High Energy Physics Group at Oklahoma State University for warm hospitality.
Partial support from Shota Rustaveli National Science Foundation is kindly acknowledged.

\appendix

\renewcommand{\theequation}{A.\arabic{equation}}\setcounter{equation}{0}


\section{Breaking of $U(1)_F$}\la{apA}
\la{appA}

In this appendix we discuss the breaking of $U(1)_F$ gauge symmetry and show that desired VEVs for the flavon fields can be
generated. As was mentioned in the text of the paper, the minimal setup of the charged flavon superfields, which we consider is
$X$ and $\bar X$ with $U(1)_F$ charges given in Eq. (\ref{XbarX-ch}). Since we are dealing with Abelian flavor symmetry, in general the
Fayet-Iliopoulos (FI) term is allowed and we will include it in our consideration. It has the form $\xi \int d^4\te V_{U(1)_F}$, where
$\xi $ is parameter with dimension of mass squire. This FI term together with standard $D$-term Lagrangian couplings,  for $V_{U(1)_F}$'s  auxiliary component
give:
\beq
D_{U(1)_F}=\xi -\bt |X|^2+\bt |\bar X|^2~.
\la{D-U1}
\eeq
Moreover, in order to fix all VEVs we need to have some superpotential couplings. For this purpose we introduce the superfield $S$
which is neutral ($Q(S)=0$) under $U(1)_F$. The most general renormalizable superpotential involving $X$, $\bar X$ and $S$ will have the form
\beq
W=\lam S(X\bar X-\mu^2)+\fr{1}{2}m_SS^2+\fr{1}{3}\si S^3~,
\la{W-flavon}
\eeq
where $\mu $ and $m_S$ are some mass parameters, while $\lam $ and $\si $ are dimensionless couplings. From (\ref{W-flavon}), for $F$-components we
derive
\beq
-F_S^*=\lam(X\bar X-\mu^2)+m_SS+\si S^2~,~~~~F_X^*=-\lam S\bar X~,~~~F_{\bar X}^*=-\lam SX~.
\la{Fs}
\eeq
In the unbroken SUSY limit  $D$ and $F$-terms should satisfy
$F_S=F_X=F_{\bar X}=D_{U(1)_F}=0$, which using (\ref{D-U1}) and (\ref{Fs}) gives
\beq
|X|^2 -|\bar X|^2=\xi/\bt ~,~~~~X\bar X=\mu^2~,~~~~S=0~.
\la{cond}
\eeq
These give non-zero VEVs for $X$ and $\bar X$ fields:
\beq
|X|=\fr{1}{\sqrt{2}}\l \fr{\xi }{\bt}+\sqrt{\fr{\xi^2}{\bt^2}+4|\mu |^2}~\r^{1/2}~,~~~
|\bar X|=\sqrt{2}|\mu|^2\l \fr{\xi }{\bt}+\sqrt{\fr{\xi^2}{\bt^2}+4|\mu |^2}~\r^{-1/2}~.
\la{solXXbar}
\eeq
From (\ref{solXXbar}) we see that $|X|$ and $|\bar X|$ have different values. It is interesting to consider two limiting cases:
\begin{eqnarray}\label{pot}
{\bf a)}:&&  ~~\xi/\bt <0~,~~~|\mu |^2\ll -\xi/\bt ~, \nonumber \\
&&\hs{0.5cm}|X|\simeq \fr{|\mu |^2}{\sq{-\xi/\bt }}~,~~~|\bar X|\simeq \sq{-\xi/\bt }~,~~~|X|\ll |\bar X|~, \nonumber \\
{\bf b)}:&&  ~~\xi/\bt >0~,~~~|\mu |^2\ll \xi/\bt ~,  \nonumber \\
&& \hs{0.5cm}|X|\simeq  \sq{\xi/\bt }~,~~~|\bar X|\simeq  \fr{|\mu |^2}{\sq{\xi/\bt }} ~,~~~|X|\gg |\bar X|~.
\la{sols}
\end{eqnarray}
Thus, with notations of Eq. (\ref{eps-epsbar}), case {\bf a)} gives $\ep \ll \bar{\ep}$, while in case {\bf b)} we have $\ep \gg \bar{\ep}$.
When the scales satisfy relation $\fr{\xi}{\bt}\sim |\mu|^2$, Eq. (\ref{solXXbar}) gives $\ep \sim \bar{\ep}$.
Note that with solution (\ref{solXXbar}) and $\lan S\ran =0$, all states coming from the superfields $X$, $\bar X$ and $S$ get masses.

Including soft SUSY breaking terms in the potential, VEVs of the fields will be slightly shifted. In particular, with soft mass squires $m_X^2$
and $m_{\bar X}^2$ for the fields $X$ and $\bar X$ respectively, one can readily check that their VEVs are shifted in such a way that
$D_{U(1)_F}\simeq 2(m_X^2-m_{\bar X}^2)/\tl{g}^2$ ($\tl{g}$ is $U(1)_F$'s coupling constant). As discussed in the end of Sect. \ref{sect:models},
this would have impact on flavor violating processes. On the other hand, within minimal SUGRA scenario, the universality
$m_X^2=m_{\bar X}^2$ insures that $D_{U(1)_F}=0$.

\bibliographystyle{unsrt}

\end{document}